\documentclass[12pt]{article}  
\usepackage{graphicx}
\usepackage[centertags]{amsmath}
\usepackage[english]{babel}
\usepackage{amsfonts}
\usepackage{amssymb}
\usepackage{amsthm}
\usepackage{fancyhdr}  
\usepackage{setspace}   
\usepackage{caption}

\usepackage{amsmath}
\usepackage{algorithm}
\usepackage[noend]{algpseudocode}
\usepackage{multirow}

\title{\textbf{STAD Research Report 01\_2014} \\
\vspace{12pt}
Accurate algorithms for identifying the median ranking when dealing with weak and partial rankings under the Kemeny axiomatic approach.}

\author{Sonia Amodio**, Antonio D'Ambrosio**,Roberta Siciliano*\\
\\**Department of Industrial Engineering
\\***Department of Economics and Statistics. 
\\University of Naples Federico II\\ 
e-mails: \{sonia.amodio,antdambr,roberta\}@unina.it}

\date{April 18, 2014}

\begin{document}
\maketitle


\noindent \textbf{Abstract}.
Preference rankings virtually appear in all field of science (political sciences, behavioral sciences, machine learning, decision making and so on). The well-know social choice problem consists in trying to find a reasonable procedure to use the aggregate preferences expressed by subjects (usually called judges) to reach a collective decision. This problem turns out to be equivalent to the problem of estimating the consensus (central) ranking from data that is known to be a  NP-hard Problem. Emond and Mason in 2002 proposed a branch and bound algorithm to calculate the consensus ranking given $n$ rankings expressed on $m$ objects. Depending on the complexity of the problem, there can be multiple solutions and then the consensus ranking may be not unique. We propose a new algorithm to find the consensus ranking that is equivalent to Emond and Mason's algorithm in terms of at least one of the solutions reached, but permits a really remarkable saving in computational time.\\


\section{Introduction}
\label{sec:1}
\begin{figure}[!h]
          \centering
                \includegraphics[width=0.85\textwidth]{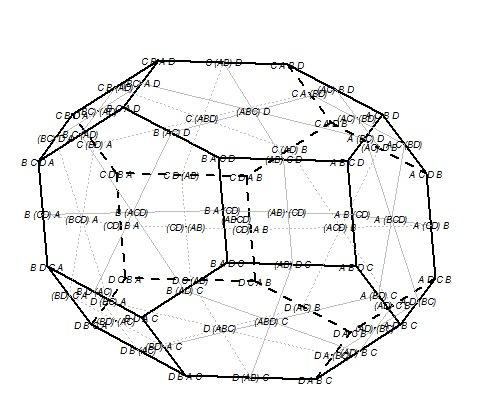}
                \caption{Permutation polytope for all full and weak ranking for four objects. For every ranking the correspondent ordering is shown.}
                \label{fig:polytope}
\end{figure}

The consensus ranking problem, also known as social choice problem, arises any time $n$ subjects (or \textit{judges}) are asked to express their preferences on a set of $m$ objects. These objects are placed in order by each subject (where $1$ represents the best and $m$ the worst) without any attempt to describe how much one differs from the others or whether any of the alternatives is good or acceptable.
Every independent observation is a permutation of $m$ distinct positive integer numbers. To be more specific, when the subject assigns the integer values from 1 to $m$ to all the $m$ items we have a \textit{complete} (or full) ranking. Whenever instead the judge fails to distinguish between two or more items and assigns to them the same integer number (expressing indifference to the relative order of this set of items), we deal with \textit{tied} (or weak) rankings. Moreover we have a \textit{partial} ranking when judges are asked to rank a subset of the entire set of objects (e.g. pick the three most favourite items out of a set of five) \cite{marden1996analyzing,AmbrHeis}.
Rankings are by nature peculiar data in the sense that the sample space of $m$ objects can be only visualized in a $(m-1)$-dimensional hyperplane by a discrete structure that is called the \textit{permutation polytope}, $S_m$. A polytope is a convex hull of a finite set of points in $\mathbb{R}^m$ \cite{thompson1993generalized,heiser2004geometric}. For example the space considering 4 objects with all possible ties is a truncated octahedron that can be visualized in Figure \ref{fig:polytope} \cite{heiser2013clustering}. As we already pointed out, the permutation polytope is inscribed in a $(m-1)$-dimensional subspace, hence, for $m > 4$, such structures are impossible to visualize.
\noindent The permutation polytope is the natural space for ranking data. To define it no data are required, it is completely determined by the number
of items involved in the preference choice; data add only information on which rankings occur and with what frequency they occur. This space is discrete and finite. It is characterized by symmetries and it is endowed with a graphical metric.\\
The problem of combining rankings to obtain a ranking representative of the group has been studied by numerous researchers in several areas, e.g. voting systems, economics, machine learning, psychology, political sciences, for more than two centuries. In the framework of distance-based models for rankings, searching for consensus ranking is a very important step in modeling the ranking process \cite{marden1996analyzing}. These models are usually exponential family models \cite{Diaconis1988} and they are completely specified by two parameters, a dispersion parameter and a consensus (central) ranking.
Maximum likelihood estimates of the dispersion parameter assume the knowledge of the central ranking.
When the consensus ranking is not known it should be estimated. Unfortunately, even if there are close formulas for this estimation they are not feasible because of the complexity of the problem \cite{critchlow1985lectu,Fligner1986,Fligner1988,Diaconis1988,Critchlow1991}.
Several methods to aggregate individual preference rankings have been proposed since the works of \cite{Borda1781}, \cite{de1785essai}, \cite{black1958theory}, \cite{arrow1951social}, \cite{Goodman1952}, \cite{Coombs1964}, \cite{Davis1972}, \cite{bogart1973preference}, \cite{Cook1978}, \cite{barthelemy1981median}, \cite{Emond2002} and \cite{meila2012consensus}.\\
In this paper, we propose two heuristic algorithms called QUICK and FAST to derive the consensus ranking from the aggregation of individual preferences within the Kemeny and Snell axiomatic framework. Both algorithms can be viewed as alternatives to the branch-and-bound algorithm by Emond and Mason. The BB algorithm turns out to be a time consuming procedure when the number of objects is high and especially when the internal degree of consensus present in the data is weak.
Both QUICK and FAST algorithms can deal with complete and tied rankings as well as with incomplete (or partial) rankings.
As a matter of fact, the QUICK algorithm is the building block of the FAST algorithm. Both provide savings in computational time, but the FAST algorithm is more accurate because it finds more than one of the solutions found by the BB algorithm and it can also easily deal with problems characterized by a large number of objects to be ranked and weak and partial rankings and/or a low degree of internal consensus. On the other hand, the QUICK algorithm turns out to be really useful when the number of objects is limited because it returns one of the solutions found by the BB, or a really close solution, in a considerably small amount of time.\\
\noindent The rest of the paper is organized as follow.
In Section 2 we briefly present some of the proposed approaches to aggregate preference rankings and derive a consensus. In Section 3 we describe the branch-and-bound algorithm by Emond and Mason. Section $4$ is devoted to describe the proposed algorithms, then in sections $5$ and $6$ we present a simulation study and applications on real data to evaluate both the accuracy and the efficiency of our proposal. Concluding remarks are then found in section $7$.\\
\section{Finding the consensus ranking, some approaches}
The term consensus ranking is a generic name for any ranking that summarizes a set of individual rankings. There exist two broad classes of approaches to aggregate preference rankings in order to find a consensus \cite{cook2006distance} :
\begin{itemize}
\item \textit{ad hoc methods}, which can be divided into elimination (for example the American system method, the pairwise majority rule, etc.) and non-elimination (for example Borda's methods of marks (1781), Condorcet's method (1785), etc.);
\item \textit{distance-based models}, according to which it's necessary to define a distance of the desired consensus from the individual rankings.
\end{itemize}
A more detailed description of both these approaches can be found in \cite{cook2006distance}. \\
\noindent How to aggregate subjects preferences to create a consensus is a problem that goes back to 1781 when Borda formulated the method of marks (also known as \textit{Borda's count}) for determining the winner in elections with more than 2 candidates. This method is quite simple and it is based on calculating the total rank for each alternative.
For example, if we consider the rankings in Table \ref{tab:borda}
\begin{table}[!h]
\centering
\caption{Example data to illustrate Borda's method of marks.}
\label{tab:borda}
\small{
\begin{tabular}{l rrr}
  \hline
& \multicolumn{3}{c}{Alternatives}\\
\hline
\# voters & A & B & C \\
  \hline
  12 & 2 & 1 & 3 \\
  5 & 1 & 2 & 3 \\
  7 & 3 & 2 & 1 \\
   \hline
\end{tabular}
}
\end{table}
\noindent the total rank for each alternative is given by:
\begin{itemize}
\item $A = 12 \times 2 + 5 \times 1 + 7 \times 3 = 50$,
\item $B = 12 \times  1 + 5 \times 2 + 7 \times 2 = 36$,
\item $C = 12 \times 3 + 5 \times 3 + 7 \times 1 =  58$,
\end{itemize}
\noindent resulting in the consensus  $(BAC)$.
Borda's method of marks was criticized by Condorcet, which proposed to use the majority rule on all the pairwise comparisons between alternatives.
Condorcet's solution for the rankings reported in Table \ref{tab:borda} can be obtained by calculating the support obtained by every pairwise comparison between options, reported in Table \ref{tab:condorcet}.
\begin{table}[!h]
\centering
\caption{Support table to illustrate Condorcet's method on example data}
\label{tab:condorcet}
\small{
\begin{tabular}{cccc}
  \hline
 & A & B & C\\
\hline
A &  - & 5 & 17  \\
B &  19 & - & 17 \\
C &  7 & 7 & -  \\
\hline
\end{tabular}
}
\end{table}
\noindent From Table \ref{tab:condorcet} we can deduce that $B \succ A$, $B \succ C$ and $A \succ C$, resulting also in the consensus ranking $(BAC)$.
In applying this method, unfortunately, one problem can be encountered, i.e. if intransitive preferences occur the simple majority procedure breaks down (\textit{paradox of voting} \cite{arrow1951social}, according to which a set of transitive preferences can generate a global intransitive preference as group preference).\\
In the last century the rank aggregation problem has been approached from a statistical perspective. \cite{kendall1938new} was the first to propose a method to aggregate input rankings to find a consensus. He studied the consensus problem as a problem of estimation and he proposed to rank items according to the mean of the ranks assigned, thus proposing a method equivalent to Borda's one.
Moreover he suggested to consider the Spearman rank correlation coefficient $\rho$, that, given two preference rankings $R$ and $R^*$, is defined as:
\begin{equation}
\rho = 1 - \frac{6\sum_{i=1}^{n}d_i^2}{n^3-n},
\end{equation}
where $d_i^2(R, R^*) = \sum_{j=1}^{m}(R_j - R^*_j)^2$ is the squared difference between rankings $R$ and $R^*$ \cite[page 8]{kendall1948rank}.
The Spearman's $\rho$ is equivalent to the product moment correlation coefficient and it treats rankings as they are scores summing the square of ranked differences.\\
\cite{kendall1938new} proposed his own correlation coefficient, named after him as Kendall $\tau$, by introducing the concept of \textit{ranking matrices}. The ranking matrix associated with the ranking $R_i$ of $m$ objects, is a $m \times m$ matrix $\{a_{ij}\}$ whose elements are defined as:
\begin{equation}
a_{ij}=\left \{ \begin{array}{l l}
1 & \textrm{ if object $i$ is ranked ahead of object $j$ } \\
-1 & \textrm{ if object $i$ is ranked behind object $j$ } \\
0 & \textrm{ if the objects are tied, or if $i=j$ }
\end{array}\right.
\end{equation}
The Kendall correlation coefficient $\tau$ between two rankings, $R$, with score matrix $\{a_{ij} \}$, and $R^*$,
with score matrix $\{b_{ij}\}$, can be then defined as the generalized correlation coefficient:
\begin{equation}
\tau(R,R^*) = \frac{\sum_{i=1}^m {\sum_{j=1}^m{a_{ij}b_{ij}}}}{\sqrt{\sum_{i=1}^m {\sum_{j=1}^m{a_{ij}^2}} \sum_{i=1}^m {\sum_{j=1}^m{b_{ij}^2}}}} \hspace{3mm}.
\end{equation}
\noindent In the same period \cite{kemeny1959mathematics} and \cite{kemeny1962mathematical} proposed and proved an axiomatic approach to find a unique distance measure for rankings and define a consensus ranking. They introduced four axioms, reported in Table \ref{KS_ax}, that should apply to any distance measure between two rankings.
\begin{table}[!h]
\caption{Kemeny and Snell axioms}
\label{KS_ax}
\small
\rule{1\textwidth}{0.5mm}
\begin{enumerate}
\item Axiom 1: $d(R_1, R_2)$ satisfies the three standard properties of a metric (or distance):
\begin{enumerate}
\item  \textit{Positivity}, $d(R_1,R_2)\geq 0$, with equality if and only if $R_1\equiv R_2$.
\item  \textit{Symmetry}, $d(R_1,R_2) = d(R_2,R_1)$.
\item  \textit{Triangular inequality}, $d(R_1,R_3) \leq d(R_1,R_2) + d(R_2,R_3)$ for any three rankings $R_1$, $R_2$, $R_3$, with equality holding if and only if ranking $R_2$ is between $R_1$ and $R_3$.
\end{enumerate}
\item Axiom 2 : Invariance \\
  $d(R_1,R_2) = d(R_1',R_2')$, where $R_1'$ and $R_2'$ result from $R_1$ and $R_2$ respectively by the same permutation of the alternatives.
\item Axiom 3: Consistency in measurement \\
  If two rankings $R_1$ and $R_2$ agree except for a set $S$ of $k$ elements, which is a segment of both, then $d(R_1,R_2)$ may be computed as if these $k$ objects were the only objects being ranked.
\item Axiom 4: Scaling \\
  The minimum positive distance is 1.
\end{enumerate}
\rule{1\textwidth}{0.5mm}
\end{table}
\noindent They also proved the existence of a distance metric that satisfies all these axioms, known as \textit{Kemeny distance}, and its uniqueness.
By using the score matrices as defined by Kendall, Kemeny's distance between two rankings $R$ (with score matrix $\{a_{ij} \}$) and $R^*$
(with score matrix $\{b_{ij}\}$) is defined as:
\begin{equation}
d_{Kem}(R, R^*) = \frac{1}{2} \sum_{i=1}^m{\sum_{j=1}^m{|a_{ij} - b_{ij}|}} \hspace{3mm}.
\end{equation}
Kemeny and Snell then suggested the idea to use this distance function to define the median ranking as a specific definition of consensus ranking. According to their definition, the median ranking is the point in the ranking space that shows the best agreement with the set of input rankings. More formally, given a set of $n$ independent input rankings $\{R_i\}_{ i=1}^n$, the median ranking $\hat{S}$ is the point (or the points) for which $\sum_{i=1}^n d(R_i, S)$ is a minimum. \\
\noindent Following the Kemeny and Snell approach, the research of the median ranking requires searching the space of all possible rankings of $m$ object.
Given a set of $n$ independent input rankings the problem consists in finding the ranking $\hat{S}$ that best represents the combined preferences of the judges. This is a NP-hard problem. When we have $m$ objects, there are $m!$ possible complete rankings. In case we deal with tied rankings, the analysis is more complex as, by including ties, the number of possible rankings approximates $\frac{1}{2} (\frac{1}{ln(2)} )^{m+1} m!$
\cite{gross}. In other words, the complexity of the search of the median ranking is entirely determined by the number of objects to be ranked. \\
\cite{bogart1973preference, bogart1975preference} generalized the Kemeny and Snell approach by considering both transitive and intransitive preferences.
\cite{Cook1976} proposed a branch-and-bound algorithm to determine the median ranking out of a set of $n$ independent preference rankings deriving a solution by adjacent pairwise optimal rankings. Emond and Mason (2002) pointed out that Cook and Saipe's method does not guarantee that all solutions are found and in some examples local optima were encountered.
\cite{cook1997general} proposed a general representation of distance-based consensus with the aim of associating a value to rank positions and developed models for deriving a  consensus.
\cite{Cook2007} presented a branch-and-bound algorithm for finding the consensus ranking in presence of partial rankings, but not allowing for ties. \\
\cite{Emond2002} proposed a new rank correlation coefficient called $\tau_x$ that is equivalent to the Kemeny and Snell distance metric. They defined the score matrices in a slightly different way respect to the Kendall's representation:  $a_{ij}=1$ if object $i$ is either ranked ahead  \textbf{or tied with} object $j$, and $a_{ij}=0$ only if $i=j$. Using these score matrices, they defined their rank correlation coefficient as:
\begin{equation}
\tau_x = \frac{\sum_{i=1}^m\sum_{j=1}^m a_{ij}b_{ij}}{m(m-1)}.
\end{equation}
Note that $\tau_x$ is equivalent to Kendall's $\tau$ when ties are not allowed. By using this correlation coefficient they proposed a branch-and-bound algorithm to deal with the median ranking problem when the number of object $m$ is at most equal to 20 in a reasonable computing time.
Given $n$ weak orderings of $m$ objects, $R_1,..., R_n$, where each ordering carries a positive weight, $w_k$, median ranking $\hat{S}$ is the one (or the ones) that maximizes the weighted average correlation with the $n$ input rankings or, equivalently, is the one (or the ones) that minimizes the weighted average Kemeny distance to the $n$ input rankings,
\begin{equation}
	\textrm{max} \frac{\sum_{k=1}^{n}w_k\tau_X(S,R^{(k)})}{\sum_{k=1}^n w_k}.
\end{equation}
\noindent Indicating as $\{s_{ij}\}$ and $\{r_{ij}\}^{(k)}$ the scoring matrices for $S$ and the $k_{th}$ ordering $R$, $k=1\ldots,n$, the problem is:
\begin{equation}
\textrm{max} \sum_{k=1}^{n}w_k\left\{\sum_{i=1}^{m}\sum_{j=1}^{m}s_{ij}r_{ij}^{(k)}\right\} = \textrm{max} \sum_{i=1}^{m}\sum_{j=1}^{m}s_{ij}c_{ij},
\label{maxim}
\end{equation}
where $c_{ij}=\sum_{k=1}^{n}w_kr_{ij}^{(k)}$.
\noindent The score matrix $\{c_{ij}\}$ was called by Emond and Mason \textit{Combined Input Matrix} (CI) because it is the result of a summation of each input ranking. Defined in this way, it summarizes the rankings information in a single matrix.\\
Emond and Mason conceived a branch-and-bound algorithm to maximize equation \ref{maxim} by defining an upper limit on the value of that dot product. This limit, considering that the score matrix  $\{s_{ij}\}$ consists only of the values $1$, $0$ and $-1$, is given by the sum of the absolute values of the  elements of CI:
\[
V = \sum_{i=1}^m\sum_{j=1}^m\left|c_{ij}\right|.
\]
%
\section{Emond and Mason's branch-and-bound algorithm}
\noindent If a weak ordering of $m$ objects is given as initial solution, it is possible to compute the associated score matrix $\left\{s_{ij}\right\}$ and evaluate the value of expression \ref{maxim}. Then it is possible to define an initial penalty $P$ by subtracting this value from $V$. The problem is to search the set of all weak orderings of $m$ objects to find those with the minimum penalty.
This set can be divided into three mutually exclusive branches based on the relative position of the first two objects in the ordering represented in the initial solution, labeled as $i$ and $j$. An incremental penalty for each of the branches can be calculated, by considering the corresponding elements $c_{ij}$ and $c_{ji}$ of the CI matrix, as specified in Table \ref{penbb}.
\begin{table}[!h]
\caption{Penalty computation in the BB algorithm}
\label{penbb}
\rule{1\textwidth}{0.5mm}
Let $\delta P$ be the incremental penalty:
\begin{itemize}
\item object $i$ is preferred to object $j$ (Branch 1):\\
\indent if $c_{ij}>0$ \textit{and} $c_{ji}<0$, then $\delta P=0$\\
\indent if $c_{ij}>0$ \textit{and} $c_{ji}>0$, then $\delta P=c_{ji}$\\
\indent if $c_{ij}<0$ \textit{and} $c_{ji}>0$, then $\delta P=c_{ji}-c_{ij}$
\item object $i$ is tied with object $j$ (Branch 2):\\
\indent if $c_{ij}>0$ \textit{and} $c_{ji}<0$, then $\delta P=-c_{ji}$\\
\indent if $c_{ij}>0$ \textit{and} $c_{ji}>0$, then $\delta P=0$\\	
\indent if $c_{ij}<0$ \textit{and} $c_{ji}>0$, then $\delta P=-c_{ij}$
\item object $j$ is preferred to object $i$ (Branch 3):\\
\indent if $c_{ij}>0$ \textit{and} $c_{ji}<0$, then $\delta P=c_{ij}-c_{ji}$\\
\indent if $c_{ij}>0$ \textit{and} $c_{ji}>0$, then $\delta P=c_{ij}$\\
\indent if $c_{ij}<0$ \textit{and} $c_{ji}>0$, then $\delta P=0$
\end{itemize}
\rule{1\textwidth}{0.5mm}
\end{table}
\noindent If the incremental penalty for a branch is greater than the initial penalty, then we do not consider it any longer because all orderings in that branch will have a penalty larger than the initial one.\\
If the incremental penalty of a branch is smaller (or equal) than the initial penalty, we then consider the next object in the initial solution and create new branches by placing this object in all possible positions relative to the objects already considered.\\
The algorithm continues in an iterative way by including all other objects until all branches to be considered are checked. The BB algorithm works with complete, incomplete and partial rankings. It deals with incomplete rankings thanks to the convention that unranked objects do not add anything in forming the combined input matrix. Emond and Mason stated that the computation time needed to reach a solution(s) depends both on the inherent degree of consensus in the sample of judges and on the quality of the initial solution used to initialize the algorithm.  For an extensive discussion on the branch-and-bound algorithm we refer to \cite{Emond2000, Emond2002}.
\section{FAST and QUICK algorithms}
\noindent The first element to be evaluated in developing our algorithm is the combined input matrix. This matrix contains all the information about the rankings expressed by all the subjects and, if it is a valid score matrix, then the median ranking can be found immediately. Unfortunately such a situation rarely happens. But by evaluating the CI with more attention it is possible to identify a good candidate to be the median ranking that can be used as an input in the algorithm. Let $Q=\bold{1}$ be a vector of ones of size $m$. Let $\{c_{ij}\}$ be the $m \times m$ combined input matrix. By taking into account all the combinations of m objects, each pair of items is evaluated once by considering the two associated cells in CI. A moderately accurate first candidate to be the median ranking can be computed as follow:\\
If sign $c_{ij}=1$ \& $\textrm{sign } c_{ji}=-1$, then $Q_i=Q_i+1$;\\
If sign $c_{ij}=-1$ \& $\textrm{sign }c_{ji}=1$, then $Q_j=Q_j+1$;\\
If sign $c_{ij}=1$ \& $\textrm{sign } c_{ji}=1$, then $Q_i=Q_i+1, Q_j=Q_j+1$.\\
In this way, we obtain the updated rank vector $Q$ containing the number of times each object is preferred to the others in the pairwise comparisons. This vector is the starting point of our algorithm. The first step is to compute the score matrix, $\{q_{ij}\}$, associated with $Q$. Then we compute the associated penalty as:
\begin{equation}
P=V-\sum_{ij}c_{ij}q_{ij}
\label{pen}
\end{equation}
After this step we take into account the object in $Q$ ranked at the second position, and we evaluate equation \ref{pen} by placing that object in all possible positions relative to the object ranked ahead, including ties.
In other words, in the first step the second ranked object is placed ahead, in a tie and behind the first object, keeping all other objects fixed in the initial position. The penalty (equation \ref{pen}) is then evaluated for these three rankings and we continue by evaluating only the ranking with the lowest penalty in the successive step, updating the candidate median ranking.
Once the penalties are computed, we update the candidate by selecting the ranking that is associated with the minimum penalty. Subsequently we add the object ranked in the third position in the initial $Q$ vector, and again we compute the values of equation \ref{pen} by placing that object in all possible positions relative to the objects already ranked ahead, including all possible ties. As before, we update the candidate median ranking by selecting the one that minimizes the penalty.  We continue in this way until all the objects are processed and we reach a possible solution.\\
We use, then, the obtained solution as starting point for a new complete loop. The overall procedure is repeated again by considering also the reverse ranking of the initial $Q$ vector as candidate median ranking. The complete algorithm is summarized in Box \ref{quick}.
\begin{algorithm}
 \floatname{algorithm}{Box}
\caption{QUICK algorithm for the median ranking problem}
\label{quick}
\begin{algorithmic}
\State \textbf{input} $\{c_{ij}\}$, $Q$
\State \textbf{initialize:} fix the rank of the first ranked object in $Q$
\State \begin{itemize}
\item[\textbf{(1.)}] consider the next ranked object in $Q$
\begin{itemize}
\item[\textbf{(2.)}] evaluate eq. \ref{pen} for all the rankings obtained by placing that object in all possible positions wrt the fixed ranked objects
\item[\textbf{(3.)}] store only the ranking associated with minimum value of eq. \ref{pen}
\end{itemize}
\item[\textbf{(4).}] fix the rank of the processed object and return to step (1.) until all objects in $Q$ are processed
\end{itemize}
\State Obtain the update ranking $CR$, and repeat all previous steps by replacing $Q$ with $CR$
\State \textbf{output:} $CR=$ median ranking.
\end{algorithmic}
\end{algorithm}
\noindent Note that when we evaluate the penalty, we consider all the objects in the ranking that is considered as candidate solution. This is a fundamental difference with the original algorithm, because Emond and Mason calculate the penalty values only by considering the elements of the combined input matrix associated with the processed objects, and updating the penalty by adding up these partial values. Indeed, we never use this penalty update.\\ We call this algorithm ``QUICK'' because it is able to reach at least one solution, or a solution really close to the true one, in few seconds even when working with a huge number of objects. In our experience, by using our definition of starting point $Q$, at least one solution is found. But, sometimes, solutions were also reached with random starting points. For this reason, we decided to use the QUICK algorithm as building block of our accurate FAST algorithm for the median ranking problem, whose pseudo-code is shown in Box \ref{al1}.  Of course, our FAST algorithm is useful when the complexity of the problem is really intractable, e.g. when the number of objects to be ranked is high and the internal degree of consensus is low.
\begin{algorithm}[!h]
 \floatname{algorithm}{Box}
\caption{Accurate FAST algorithm}
\label{al1}
\begin{algorithmic}
\State \textbf{input} $\{c_{ij}\}$
\For{iter=1:maxiter}
		\If{iter=1}
		\State CR=QUICK($Q$,$\{c_{ij}\}$), with $Q$ as defined before
		\State store CR
		\Else
		\State $Q$=random permutation of $m$ objects
		\State CR=QUICK($Q$,$\{c_{ij}\}$)
		\State store CR
		\EndIf
		\EndFor
	\State \textbf{output:} CR=CR:$\tau_x$=max
\end{algorithmic}
\end{algorithm}
\noindent Among the solutions returned by the QUICK algorithm, the median rankings are those showing the highest value of the average $\tau_x$ rank correlation coefficient.
\section{Simulation study}
\begin{figure}[!h]
          \centering
                \includegraphics[width=0.85\textwidth]{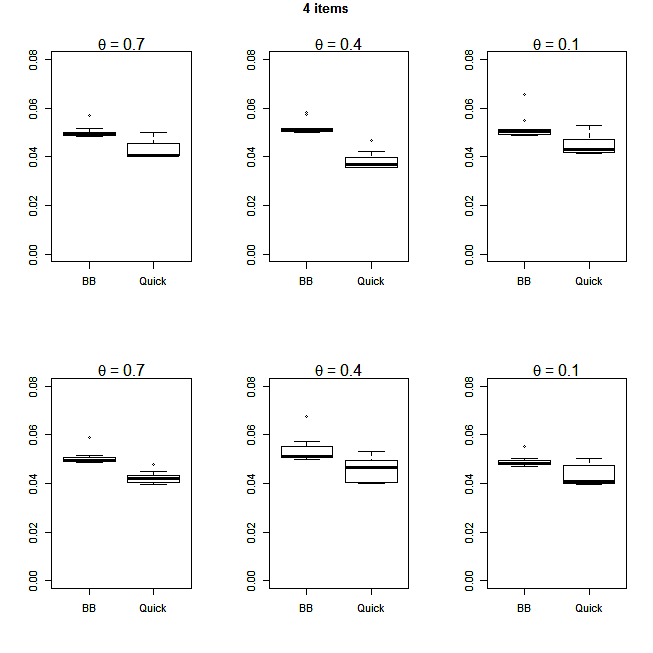}
                \caption{Working time in second. The first row of box-plots refers to complete rankings, the second row refers to tied and complete rankings}
                \label{4it}
\end{figure}

\begin{figure}[!h]
          \centering
                \includegraphics[width=0.85\textwidth]{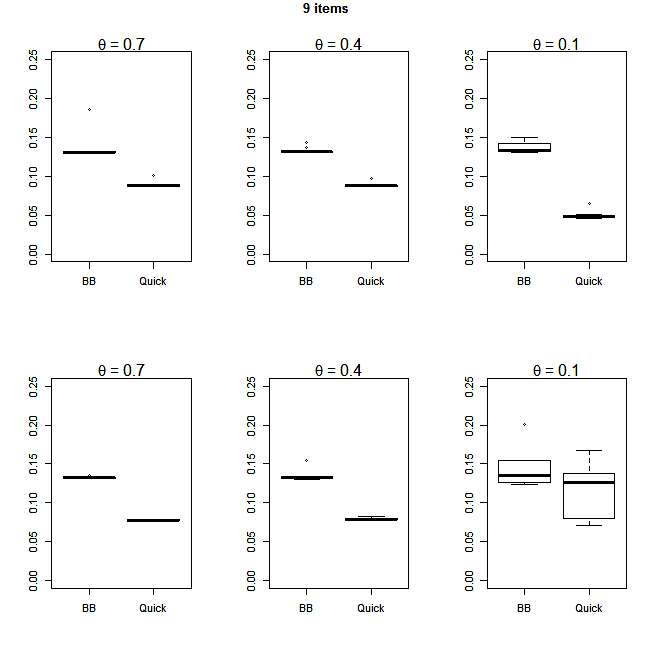}
                \caption{Working time in second. The first row of box-plots refers to complete rankings, the second row refers to tied and complete rankings}
                \label{9it}
\end{figure}

\begin{figure}[!h]
          \centering
                \includegraphics[width=0.85\textwidth]{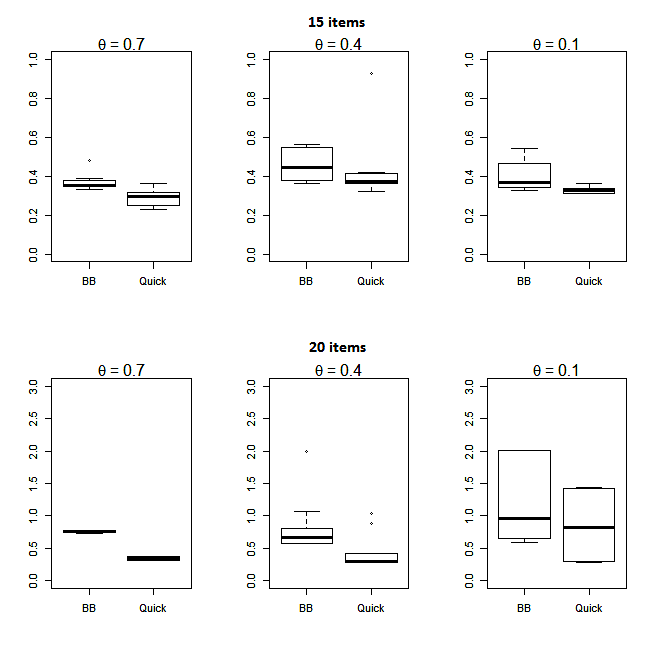}
                \caption{Working time in second.The first row of box-plots refers to complete rankings on 15 objects, the second row refers to complete rankings on 20 objects.}
                \label{1520it}
\end{figure}

We implemented the BB algorithm by Emond and Mason, as well as both the QUICK and FAST algorithms in MatLab and in R environments. The reported results are based on codes written in MatLab language. A beta version of the R \textit{ConsRank} package is available upon request to the authors, as well as the MatLab codes. Analysis were made by using a Computer Intel Core i5-3317U 1.70 GHz and 4GB of RAM.\\
To evaluate the performance of our algorithms in terms of accuracy and efficiency, we performed a simulation study. Ranking data were simulated according to a distance-based model by selecting three different levels of the dispersion parameter $\theta$, which governs the degree of consensus in the sample of rankings.
In the distance-based models framework, for a given consensus, $S$, a distance function, $d$, and some real parameter $\theta$, the density with respect to the \textit{Uniform} distribution is equal to
\[
f_\theta  \left( {a;S} \right) = C(\theta ) \exp \left( -\theta d\left( {S,a} \right) \right),
\]
where $a$ is a ranking and $C(\theta)$ is a normalizing constant. For more details on distance-based models we refer to \cite{marden1996analyzing}, \cite{Feigin1978} and \cite{Critchlow1991}.\\
The three levels chosen for $\theta$ were $0.7$, $0.4$ and $0.1$, the distance used was the Kemeny distance.
We decided to consider $4$ different levels for $m$: $4$, $9$, $15$ and $20$. In the case of $4$ and $9$ objects, we repeated the experiment both considering only complete rankings and the full space of complete and tied rankings, while in the case of $15$ and $20$ objects we decided to limit the experiment only to complete rankings sampled from a limited sub-population of size 10 millions. These sub-populations were generated from the full rankings space of $10$ objects by adding the remaining objects in such a way that they were at first ranked below, later ranked ahead, and then randomly ranked in a middle position. Sample size was always equal to $200$. Another experiment involved incomplete rankings. We chose a scheme of the type ``pick k out of m'', and precisely: pick $2$ out of $4$, pick $5$ out of $9$ and pick $10$ out of $15$. Rankings were sampled in this way: first we extracted a random number of rankings (from a minimum of $15$ to a maximum of $30$)  according to the uniform distribution by setting $\theta=0$ from the corresponding spaces, then we generated the weights from a normal distribution with means randomly generated between $10$ and $30$ and standard deviations randomly generated between $2.5$ and $9$. After we normalized the weights and multiplied them by the total sample size to have data sets approximatively of size $200$. Each experiment was repeated ten times, for globally $240$ data sets. Table \ref{fac} summarizes the experimental design.
\begin{table}[!h]
  \centering
\caption{Experimental factors by levels}
\label{fac}
\small
 \tabcolsep=0.11cm
\scalebox{0.85}{
  \begin{tabular}{|c|c|c|}
	\hline
	  Objects & Rankings & $\theta \backslash$Distribution \\
    \hline
    \multirow{6}{*}{4} & \multirow{3}{*}{Full} & 0.7\\
		 & & 0.4 \\
		 & & 0.1 \\
		 \cline{2-3}
		 & \multirow{3}{*}{Tied} & 0.7 \\
		 & & 0.4 \\
		 & & 0.1 \\
    \hline
		\multirow{6}{*}{9} & \multirow{3}{*}{Full} & 0.7\\
		 & & 0.4 \\
		 & & 0.1 \\
		\cline{2-3}
		 & \multirow{3}{*}{Tied} & 0.7 \\
		 & & 0.4 \\
		 & & 0.1 \\
    \hline
		\multirow{3}{*}{15} & \multirow{3}{*}{Full} & 0.7\\
		 & & 0.4 \\
		 & & 0.1 \\
    \hline
		\multirow{3}{*}{20} & \multirow{3}{*}{Full} & 0.7\\
		 & & 0.4 \\
		 & & 0.1 \\
    \hline
		\hline
		\multirow{2}{*}{pick 2 out of 4} & \multirow{2}{*}{Incomplete} & Normal\\
		& & Uniform \\
		\hline
		\multirow{2}{*}{pick 5 out of 9} & \multirow{2}{*}{Incomplete} & Normal\\
		& & Uniform \\
		\hline
		\multirow{2}{*}{pick 10 out of 15} & \multirow{2}{*}{Incomplete} & Normal\\
		& & Uniform \\
		\hline		
  \end{tabular}
	}
\end{table}
\noindent For each data set we ran the BB, the QUICK and the FAST algorithms. We checked the median rankings found by the three algorithms as well as the elapsed time in seconds to reach the solutions. We used the BB algorithm as benchmark to check the accuracy of our algorithms in terms of solutions. The initial solution for all the algorithms was the updated rank vector $Q$ as defined in section 4.\\
Table \ref{summ} shows in the first column a summary of the solutions reached by the BB algorithm. In the second and in the third columns respectively summary measures of the coincident solutions returned by the QUICK and FAST algorithms with respect to the ones handed back by the Emond and Mason's one are shown. Note that always both QUICK and FAST algorithms found at least one solution, and the proportion of solutions found by the FAST algorithm was always higher (or equal) to the one returned by the QUICK. There were no relevant differences among the factors of the experimental design except, as expected, that the lower $\theta$, the higher the number of solutions identified. This is due to the fact that in this particular experiment there was a moderate internal degree of consensus present in the data, even when $\theta$ was set equal to $0.1$.
\begin{table}[!h]
\caption{Summary measures of the number of solutions reached by BB algorithm and of the number of coincident solutions found by QUICK and FAST by number of objects, experiment with complete and tied rankings.}
\label{summ}
\centering
\small
\tabcolsep=0.11cm
\scalebox{0.9}{
\begin{tabular}{|llll|}
\hline
 & \multicolumn{1}{c}{BB solutions} & \multicolumn{1}{c}{QUICK} & \multicolumn{1}{c|}{FAST} \\
\hline
\multicolumn{1}{|c}{4 objects} &  &  &  \\
\hline
Mean & \multicolumn{1}{r}{1.2} & \multicolumn{1}{r}{1.1} & \multicolumn{1}{r|}{1.2} \\
Median & \multicolumn{1}{r}{1.0} & \multicolumn{1}{r}{1.0} & \multicolumn{1}{r|}{1.0} \\
Minimum & \multicolumn{1}{r}{1.0} & \multicolumn{1}{r}{1.0} & \multicolumn{1}{r|}{1.0} \\
Maximum & \multicolumn{1}{r}{3.0} & \multicolumn{1}{r}{2.0} & \multicolumn{1}{r|}{3.0} \\
\hline
\multicolumn{1}{|c}{9 objects} & \multicolumn{1}{r}{} & \multicolumn{1}{r}{} & \multicolumn{1}{r|}{} \\
\hline
Mean & \multicolumn{1}{r}{1.2} & \multicolumn{1}{r}{1.1} & \multicolumn{1}{r|}{1.2} \\
Median & \multicolumn{1}{r}{1.0} & \multicolumn{1}{r}{1.0} & \multicolumn{1}{r|}{1.0} \\
Minimum & \multicolumn{1}{r}{1.0} & \multicolumn{1}{r}{1.0} & \multicolumn{1}{r|}{1.0} \\
Maximum & \multicolumn{1}{r}{9.0} & \multicolumn{1}{r}{2.0} & \multicolumn{1}{r|}{5.0} \\
\hline
\multicolumn{1}{|c}{15 objects} & \multicolumn{1}{r}{} & \multicolumn{1}{r}{} & \multicolumn{1}{r|}{} \\
\hline
Mean & \multicolumn{1}{r}{2.6} & \multicolumn{1}{r}{1.4} & \multicolumn{1}{r|}{1.9} \\
Median & \multicolumn{1}{r}{1.0} & \multicolumn{1}{r}{1.0} & \multicolumn{1}{r|}{1.0} \\
Minimum & \multicolumn{1}{r}{1.0} & \multicolumn{1}{r}{1.0} & \multicolumn{1}{r|}{1.0} \\
Maximum & \multicolumn{1}{r}{18.0} & \multicolumn{1}{r}{3.0} & \multicolumn{1}{r|}{6.0} \\
\hline
20 objects & \multicolumn{1}{r}{} & \multicolumn{1}{r}{} & \multicolumn{1}{r|}{} \\
\hline
Mean & \multicolumn{1}{r}{2.6} & \multicolumn{1}{r}{1.2} & \multicolumn{1}{r|}{1.9} \\
Median & \multicolumn{1}{r}{1.0} & \multicolumn{1}{r}{1.0} & \multicolumn{1}{r|}{1.0} \\
Minimum & \multicolumn{1}{r}{1.0} & \multicolumn{1}{r}{1.0} & \multicolumn{1}{r|}{1.0} \\
Maximum & \multicolumn{1}{r}{9.0} & \multicolumn{1}{r}{2.0} & \multicolumn{1}{r|}{5.0} \\
\hline
\end{tabular}
}
\end{table}
\noindent Table \ref{pickprop} reports the solutions returned by the BB algorithm and the number of coincident solutions recovered by the QUICK and FAST algorithms in the experiment with incomplete rankings.
In this case, due to the sampling procedure, the internal degree of consensus in the data sets was quite poor. The experiments with $9$ and $15$ objects respectively count a maximum number of solutions equal to $31$ and $7761$. In one case the QUICK algorithm did not find one of the BB solutions, but it did not happen with the FAST algorithm. This particular case is helpful to understand why we called this algorithm ``FAST''. The BB algorithm found $25$ solutions in $24240.054$ seconds ($\sim 6.733$ hours), each one with an average $\tau_x$ equal to $0.106$. The FAST algorithm could find $6$ of the $25$ solutions in $64.932$ seconds. The two solutions found by the QUICK algorithm were found in $0.693$ seconds and were really close to be real solutions because they were characterized by an average $\tau_x$ equal to $0.104$. This was the unique case in which the QUICK algorithm did not find one of the BB solutions.
\begin{table}[!h]
\caption{Summary measures of the number of solutions reached by BB algorithm and of number of coincident solutions found by QUICK and FAST, experiment with incomplete rankings.}
\label{pickprop}
\centering
\small
\tabcolsep=0.11cm
\scalebox{0.9}{
\begin{tabular}{|llll|}
\hline
 & \multicolumn{1}{c}{BB solutions} & \multicolumn{1}{c}{QUICK} & \multicolumn{1}{c|}{FAST} \\
\hline
\multicolumn{1}{|c}{2 out of 4} &  &  &  \\
\hline
Mean & \multicolumn{1}{r}{1.5} & \multicolumn{1}{r}{1.3} & \multicolumn{1}{r|}{1.5} \\
Median & \multicolumn{1}{r}{1.0} & \multicolumn{1}{r}{1.0} & \multicolumn{1}{r|}{1.0} \\
Minimum & \multicolumn{1}{r}{1.0} & \multicolumn{1}{r}{1.0} & \multicolumn{1}{r|}{1.0} \\
Maximum & \multicolumn{1}{r}{3.0} & \multicolumn{1}{r}{2.0} & \multicolumn{1}{r|}{3.0} \\
\hline
\multicolumn{1}{|c}{5 out of 9} & \multicolumn{1}{r}{} & \multicolumn{1}{r}{} & \multicolumn{1}{r|}{} \\
\hline
Mean & \multicolumn{1}{r}{7.4} & \multicolumn{1}{r}{2.1} & \multicolumn{1}{r|}{3.7} \\
Median & \multicolumn{1}{r}{4.0} & \multicolumn{1}{r}{2.0} & \multicolumn{1}{r|}{2.5} \\
Minimum & \multicolumn{1}{r}{1.0} & \multicolumn{1}{r}{1.0} & \multicolumn{1}{r|}{1.0} \\
Maximum & \multicolumn{1}{r}{31.0} & \multicolumn{1}{r}{4.0} & \multicolumn{1}{r|}{12.0} \\
\hline
\multicolumn{1}{|c}{10 out of 15} & \multicolumn{1}{r}{} & \multicolumn{1}{r}{} & \multicolumn{1}{r|}{} \\
\hline
Mean & \multicolumn{1}{r}{451.0} & \multicolumn{1}{r}{1.6} & \multicolumn{1}{r|}{13.1} \\
Median & \multicolumn{1}{r}{8.0} & \multicolumn{1}{r}{1.5} & \multicolumn{1}{r|}{4.0} \\
Minimum & \multicolumn{1}{r}{1.0} & \multicolumn{1}{r}{\textbf{0.0}} & \multicolumn{1}{r|}{1.0} \\
Maximum & \multicolumn{1}{r}{7761.0} & \multicolumn{1}{r}{3.0} & \multicolumn{1}{r|}{102.0} \\
\hline
\end{tabular}
}
\end{table}
\noindent Figures \ref{4it}, \ref{9it} and \ref{1520it} show the distribution of working time of both BB and QUICK algorithms. We do not show the box-plots relative to the FAST algorithm because its computing time was approximately equal to the number of iterations multiplied by the computing time of the QUICK algorithm. As it can be noted, the QUICK algorithm is on average faster than the BB algorithm, and the variability of the computing time increases as the value of $\theta$ decreases.\\

\noindent Table \ref{picktime} summarizes the computing time for the experiment involving incomplete rankings. The computation time for the QUICK algorithm has not a considerable variability while, especially in the case of $15$ objects, BB computational time shows a higher variability.
\begin{table}[!h]
\caption{Summary measures of elapsed times (in seconds) for finding the solutions}
\label{picktime}
\centering
\small
\tabcolsep=0.11cm
\scalebox{0.9}{
\begin{tabular}{|llll|}
\hline
 & \multicolumn{1}{c}{BB } & \multicolumn{1}{c}{QUICK} & \multicolumn{1}{c|}{FAST} \\
\hline
\multicolumn{1}{|c}{2 out of 4} &  &  &  \\
\hline
Mean & \multicolumn{1}{r}{0.031} & \multicolumn{1}{r}{0.012} & \multicolumn{1}{r|}{0.337} \\
Median & \multicolumn{1}{r}{0.012} & \multicolumn{1}{r}{0.010} & \multicolumn{1}{r|}{0.318} \\
Minimum & \multicolumn{1}{r}{0.009} & \multicolumn{1}{r}{0.008} & \multicolumn{1}{r|}{0.261} \\
Maximum & \multicolumn{1}{r}{0.097} & \multicolumn{1}{r}{0.027} & \multicolumn{1}{r|}{0.595} \\
\hline
\multicolumn{1}{|c}{5 out of 9} & \multicolumn{1}{r}{} & \multicolumn{1}{r}{} & \multicolumn{1}{r|}{} \\
\hline
Mean & \multicolumn{1}{r}{0.282} & \multicolumn{1}{r}{0.170} & \multicolumn{1}{r|}{14.328} \\
Median & \multicolumn{1}{r}{0.287} & \multicolumn{1}{r}{0.185} & \multicolumn{1}{r|}{16.278} \\
Minimum & \multicolumn{1}{r}{0.218} & \multicolumn{1}{r}{0.063} & \multicolumn{1}{r|}{7.788} \\
Maximum & \multicolumn{1}{r}{0.378} & \multicolumn{1}{r}{0.219} & \multicolumn{1}{r|}{16.398} \\
\hline
\multicolumn{1}{|c}{10 out of 15} & \multicolumn{1}{r}{} & \multicolumn{1}{r}{} & \multicolumn{1}{r|}{} \\
\hline
Mean & \multicolumn{1}{r}{1967.438} & \multicolumn{1}{r}{0.745} & \multicolumn{1}{r|}{65.910} \\
Median & \multicolumn{1}{r}{255.663} & \multicolumn{1}{r}{0.686} & \multicolumn{1}{r|}{66.103} \\
Minimum & \multicolumn{1}{r}{0.745} & \multicolumn{1}{r}{0.660} & \multicolumn{1}{r|}{64.413} \\
Maximum & \multicolumn{1}{r}{24240.054} & \multicolumn{1}{r}{1.343} & \multicolumn{1}{r|}{68.537} \\
\hline
\end{tabular}
}
\end{table}
%
\section{Real data applications}
\noindent The first real data application is about the data reported by Emond and Mason (2000, pag. 28) which are shown in Table \ref{EMd}. The first $15$ columns represent the objects to be ranked with labels in the first row, while the last column reports the weight associated with every ranking.
\begin{table}[!h]
\caption{Emond and Mason's data}
\centering
\small
\tabcolsep=0.11cm
\scalebox{0.8}{
\begin{tabular}{rrrrrrrrrrrrrrrrr}
\hline
A & B & C & D & E & F & G & H & I & L & M & N & O & P & Q & $w_k$ \\
\hline
1 & 6 & 4 & 5 & - & 1 & 2 & 7 & 3 & 1 & 5 & 2 & 6 & 5 & 5 & 4 \\
11 & 10 & 4 & 8 & 9 & 1 & 7 & 12 & 2 & 3 & 2 & 6 & 13 & 5 & 14 & 4 \\
11 & 12 & 3 & 11 & 7 & 1 & 4 & 5 & 12 & 2 & 6 & 10 & 11 & 8 & 9 & 4 \\
2 & 4 & 3 & 3 & 11 & 8 & 10 & 9 & 6 & 10 & 5 & 1 & 5 & 7 & 5 & 5 \\
2 & 8 & 4 & 8 & 7 & 1 & 2 & 5 & 2 & 3 & 6 & 7 & 8 & - & - & 4 \\
2 & 9 & 5 & 1 & 4 & 3 & 2 & 7 & 3 & 1 & 8 & 6 & 3 & 4 & 8 & 5 \\
3 & 9 & 7 & 1 & 2 & 8 & 13 & 6 & 1 & 10 & 5 & 11 & 9 & 4 & 14 & 5 \\
4 & 2 & 9 & 1 & 3 & 12 & 6 & 10 & 13 & 14 & 11 & 9 & 7 & 8 & 5 & 5 \\
4 & 3 & 5 & 11 & 12 & 10 & 13 & 7 & 6 & 8 & 2 & 1 & 9 & 9 & 11 & 7 \\
4 & 7 & 8 & 6 & 13 & 2 & 3 & 12 & 9 & 1 & 5 & 10 & 5 & 11 & 11 & 4 \\
6 & 1 & 3 & 3 & 6 & 2 & 6 & 5 & 4 & 5 & 1 & 1 & 2 & 1 & 1 & 5 \\
6 & 10 & 14 & 5 & 7 & 1 & 8 & 3 & 2 & 3 & 4 & 11 & 13 & 12 & 9 & 4 \\
6 & 6 & 8 & 1 & 1 & 3 & 5 & 1 & 10 & 7 & 2 & 10 & 9 & 4 & 6 & 7 \\
7 & 2 & - & 1 & 2 & 10 & 5 & 3 & 9 & 8 & 6 & 7 & 7 & 6 & 4 & 5 \\
7 & 4 & 6 & 1 & 5 & 14 & 10 & 12 & 15 & 3 & 13 & 9 & 8 & 2 & 11 & 5 \\
7 & 8 & 4 & 5 & 7 & 1 & 6 & 5 & 3 & 2 & 7 & 9 & 10 & 11 & 12 & 4 \\
8 & 4 & 7 & 2 & 1 & 11 & 4 & 6 & 3 & 12 & 6 & 10 & 13 & 5 & 9 & 7 \\
9 & 8 & 7 & 6 & 3 & 4 & - & 2 & 5 & 1 & 3 & 7 & 6 & 4 & 6 & 7 \\
- & - & 3 & 1 & 1 & 5 & 5 & 4 & 5 & 2 & 4 & 2 & 6 & 7 & 8 & 7 \\
- & - & 4 & 7 & 2 & 10 & 11 & 5 & 8 & 8 & 9 & 1 & 2 & 3 & 6 & 7 \\
- & - & 5 & 6 & 12 & 9 & 10 & 8 & 2 & 11 & 1 & 4 & 7 & 2 & 3 & 7 \\
\hline
\end{tabular}
}
\label{EMd}
\end{table}
\noindent By using the BB algorithm we obtained exactly the following solutions (as also reported by Emond and Mason, 2000, page 29), with an average $\tau_x$ equal to $0.166$:
\begin{enumerate}
\small
\item $<$\textit{D L (E-M) (A-B) I P (C-N) H F G (O-Q)}$>$
\item $<$\textit{D L (E-M) (A-B-P) (C-N) I H F G (O-Q)}$>$
\item $<$\textit{D L (E-M) (B-P) A (C-N) I H F G (O-Q)}$>$
\end{enumerate}
Computing time was equal to $5113.608$ seconds. We ran the QUICK algorithm on these data obtaining solution number $3$ in a computing time of $0.155$ seconds. Then we ran our FAST algorithm with $100$ starting points, obtaining exactly all solutions with a computing time of $12.627$ seconds.\\
\noindent The second data set used to compare the computing time of the algorithms is the famous data set about voters for the 1980 election of American Psychological Association president \cite{Diaconis1988,murphy2003mixtures}. This data set contains the rankings expressed by 15,449 psychologists on five candidates: $A =$ Bevan, $B =$ Iscoe, $C =$ Kiesler, $D =$ Siegle and $E =$ Wriths. Of these rankings only 5,738 are complete, while the remaining are partial rankings.
\begin{table}[!h]
\caption{Median ranking on APA data set}
\label{APAres}
\centering
\tabcolsep=0.11cm
\scalebox{1}{
\begin{tabular}{ccrr}
\hline
Algorithm & solution & elapsed time & replications\\
\hline
BB & $<$C A E D B$>$ & 1.033 & - \\
QUICK & $<$C A E D B$>$ & 0.764 & - \\
FAST & $<$C A E D B$>$ & 27.814 & 50 \\
\hline
\hline
\end{tabular}
}
\end{table}
\noindent As shown in Table \ref{APAres} all the algorithms reached the same unique solution characterized by an average $\tau_x$ equal to 0.023.\\
The third data set used is known as the Sports data set and it comes from Louis Roussos \cite{marden1996analyzing}. In this data $130$ students of the University of Illinois were asked to rank seven sports according to their preference of participating in. The sports considered were: A $=$ baseball, B $=$ football, C $=$ basketball, D $=$ tennis, E $=$ cycling, F $=$ swimming and G $=$ jogging. Also in this case there is a unique solution, and the results are reported in Table \ref{Sportsres}.
\begin{table}[!h]
\caption{Median ranking on Sports data set}
\label{Sportsres}
\centering
\tabcolsep=0.11cm
\scalebox{0.95}{
\begin{tabular}{ccrr}
\hline
Algorithm & solution & elapsed time & replications\\
\hline
BB & $<$E F C A D B G$>$ & 0.076 & - \\
QUICK & $<$E F C A D B G$>$ & 0.084 & - \\
FAST & $<$E F C A D B G$>$ & 3.592 & 50 \\
\hline
\hline
\end{tabular}
}
\end{table}
\noindent Also in this case all the algorithms reach the same unique solution characterized by an average $\tau_x$ of 0.428, as reported in Table \ref{Sportsres}. \\
To test the ability of our algorithms to deal with rankings with a large number of objects the forth data set is a random subset of the rankings collected by \cite{OLearyMorgan2010} on the $50$ American States.
The number of items (the number of American States) is equal to $50$, and the number of rankings is equal to 104.
These data concern rankings of the 50 American States on three particular aspects: socio-demographic characteristics (as  population in 2008, GPD per capita, median household income, total expenditures, etc.), health care expenditures (as per capita hospital expenditures, \% of people covered by health insurance,  \% of people covered by employment base insurance, etc.) and crime statistics (as crime rate, number of arrests, murder rate, etc.).
It was unfeasible to run Emond and Mason's algorithm on this data. The orderings corresponding to the three solutions found by the FAST algorithm, characterized by an average $\tau_x$ equal to 0.298, are reported in Table \ref{Cons_am}. These solutions were obtained in $1177.274$ seconds ($\sim 19$ minutes) with $1000$ iterations. The QUICK algorithm found 1 solution (solution $2$ in Table \ref{Cons_am}) in $16.384$ seconds.
%
\begin{table}[!h]
\centering
\caption{Median ranking found by FAST algorithm, American states data}
\label{Cons_am}
\small
\tabcolsep=0.11cm
\scalebox{0.7}{
\begin{tabular}{lrrr}
\hline
&  solution 1 &  solution 2 &  solution 3 \\
\hline
1 & CA & CA & CA \\
2 & NY & NY & NY \\
3 & FL & FL & FL \\
4 & MD & MD & MD \\
5 & LA & LA & LA \\
6 & NM & NM & NM \\
7 & DE & TX & DE \\
8 & TX & IL & TX \\
9 & IL & DE & IL \\
10 & PA & PA & PA \\
11 & MI & MI & MI \\
12 & GA & GA & GA \\
13 & NC & NC & NC \\
14 & NJ & NJ & NJ \\
15 & MA & MA & MA \\
16 & WA & WA & WA \\
17 & OH & OH & OH \\
18 & VA & VA & VA \\
19 & TN & TN & TN \\
20 & NV & NV & NV \\
21 & AZ & AZ & AZ \\
22 & MO & MO & MO \\
23 & IN & IN & IN \\
24 & AK & AK & AK \\
25 & WI & WI & WI \\
26 & CO & CO & CO \\
27 & CT & CT & CT \\
28 & MN & MN & MN \\
29 & AL & AL & AL \\
30 & SC & SC & SC \\
31 & OR & OR & OR \\
32 & OK & OK & OK \\
33 & MS & MS & KY \\
34 & AR & AR & MS \\
35 & HI & HI & AR \\
36 & KY & KY & HI \\
37 & (KS - RI) & (KS - RI) & (KS - RI) \\
39 & UT & UT & UT \\
40 & (IA - NE) & (IA - NE) & (IA - NE) \\
42 & WY & WY & WY \\
43 & WV & WV & WV \\
44 & ID & ID & ID \\
45 & ME & ME & ME \\
46 & MT & MT & MT \\
47 & NH & NH & NH \\
48 & SD & SD & SD \\
49 & VT & VT & VT \\
50 & ND & ND & ND \\
$\tau_x$ & 0.298 & 0.298 & 0.298 \\
\hline
\end{tabular}
}
\end{table}
\section{Concluding remarks}
\noindent In this paper we proposed two accurate algorithms (the QUICK and the FAST) to solve the problem of identifying the median ranking in situations involving full, weak and partial ranking. Our approach lies into the Kemeny and Snell theoretical framework. Our algorithms can be considered as an alternative to branch-and-bound algorithm proposed by Emond and Mason (2002). The BB algorithm results to be a time demanding procedure when the number of objects is high especially when the degree of internal consensus in the data is weak.
Our approach is heuristic and, thus, it does not return all the possible solutions that can be found by an exhaustive search (as in the BB algorithm).
For this reason it may happen that the QUICK does not reach a solution being stuck in a local optimum, however even if this happen the FAST, by repeatedly running the QUICK algorithm with random permutations of the $m$ items, in our experiments, always identifies a global optimum.
Nevertheless, finding all the solutions in presence of multiple median rankings could not always be the final goal of the analysis, especially considering that the returned solutions are mutually coherent since they present the same value of the average $\tau_x$.
We illustrated the performance of both these algorithms in terms of accuracy and computational efficiency via simulated and real data sets.
As shown by the results of the simulation studies, when the number of objects is smaller than 15, the FAST algorithm on average recovers all the solutions handed back by the BB algorithm. On the other hand, when the number of objects is equal or higher than 15 the FAST recovers on average the $70\%$ of the BB solutions. The QUICK algorithm always finds at least one of the solutions in a sensibly lower amount of time with respect to the BB
algorithm. When dealing with partial rankings and a weak internal degree of consensus, the FAST algorithm again shows a good performance. Indeed, even if it does not return all the BB solutions, it always returns more than one solution in a limited amount of time. In this case, the QUICK also finds at least one BB solution in a considerably shorter time. Moreover, as can be noted from the real data analysis, when the number of objects is smaller than 20, the QUICK again always finds one of the BB solutions in a shorter period of time respect to the BB algorithm. If the number of objects is greater than 20, as in the case of the 50 American States data set, Emond and Masons algorithm is unfeasible, while the FAST finds three solution in less than 20 minutes.
To some extent, the impact of the result of our proposal can be compared to that one obtained by \cite{mola1997fast} in the field of classification and regression trees \cite{breiman1984classification}. As an example, \cite{siciliano2000multivariate}, \cite{Ambrosio2007} and \cite{Ambrosio2012} considered the FAST algorithm to speed up the splitting procedure in tree growing that proved to be effective respectively to deal with huge and complex data sets as well as to improve the computational cost of using ensemble methods and finally to accelerate the missing data imputation within the statistical learning paradigm.

\begin{footnotesize}
\section*{Acknowledgments} Work by Sonia Amodio was supported by Progetto Innosystem, POR Campania FSE 2007/2013, CUP B25B09000070009. 
\end{footnotesize}




\end{document}